\documentclass{article}
\usepackage{smc2026}
\usepackage[caption=false, font=footnotesize]{subfig}
\usepackage{subcaption}
\usepackage{paralist}
\usepackage[figure,table]{hypcap}
\usepackage{booktabs}
\usepackage{pgfplots}
\usepgfplotslibrary{polar}
\pgfplotsset{compat=1.18}

\usepgfplotslibrary{fillbetween}

\usepackage{stmaryrd}
\usepackage{amsmath}

\usepackage{array}
\usepackage{multirow}

\usepackage{array}
\newcolumntype{P}[1]{>{\centering\arraybackslash}p{#1}}

\usepackage[english]{babel}

\usepackage{alphabeta}

\def\papertitle{Predicting Timbre Traits for Interpretable Assessment of Musical Sound Synthesizers}

\author[1]{\mbox{\firstname{Théo}\lastname{Chasle Cauchy}\email{theo.chasle-cauchy@ls2n.fr}}}
\author[2]{\mbox{\firstname{Modan}\lastname{Tailleur}}}

\author[3]{\mbox{\firstname{Lindsey}\lastname{Reymore}}}
\author[4]{\mbox{\firstname{Fanny}\lastname{Roche}}}
\author[5]{\mbox{\firstname{Mathieu}\lastname{Lagrange}}}

\affil[1, 2, 5]{\institution{Nantes Université, École Centrale Nantes, CNRS, LS2N, UMR 6004}\postcode{F-44000}\city{Nantes}\country{France}\affiliationtype{labunit}}
\affil[3]{\institution{Arizona State University, School of Music, Dance and Theatre}\city{Tempe, AZ}\country{USA}\affiliationtype{University}}
\affil[4]{\institution{Arturia}\city{Montbonnot Saint-Martin}\country{France}\affiliationtype{Company}}

\completesetup

\title{\papertitle}
\begin{document}
	\capstartfalse
	\maketitle
	\capstarttrue

	\begin{abstract}
        Measuring neural audio synthesizers' performance is now routinely conducted using distribution based metrics such as the Fréchet Audio Distance (FAD). Although this metric can be correlated with human perception, it offers limited interpretability beyond ranking different approaches. In this paper, we introduce a deep neural timbre trait predictor composed of a pretrained audio neural embedding (CLAP), and a shallow learnable component. The latter is trained using the RWC musical instrument database and human judgments of 20 timbre descriptions (e.g., woody, percussive, rumbling, etc.) for 31 instruments. The resulting model shows strong correlation with average human ratings (\textit{r} = 0.66, \textit{p} $<$ 0.001). 
        
        We then demonstrate the benefit of this predictor for evaluating the performance of TokenSynth, a neural sound synthesizer. First, the Mean Absolute Error (MAE) computed over the set of generated sounds under different conditioning modalities of the model provides the same ranking as a FAD computed with the RWC database as a reference, suggesting that the proposed predictors are able to provide equivalent information on a distributional basis. Second, because the model is able to qualitatively analyze isolated sounds, we can determine which generated sounds could be improved and identify specific timbral dimensions that need adjustment.
	\end{abstract}

	\section{Introduction}\label{sec:introduction}

      \begin{figure*}[ht!]
        \centering
        \includegraphics[width=0.88\textwidth]{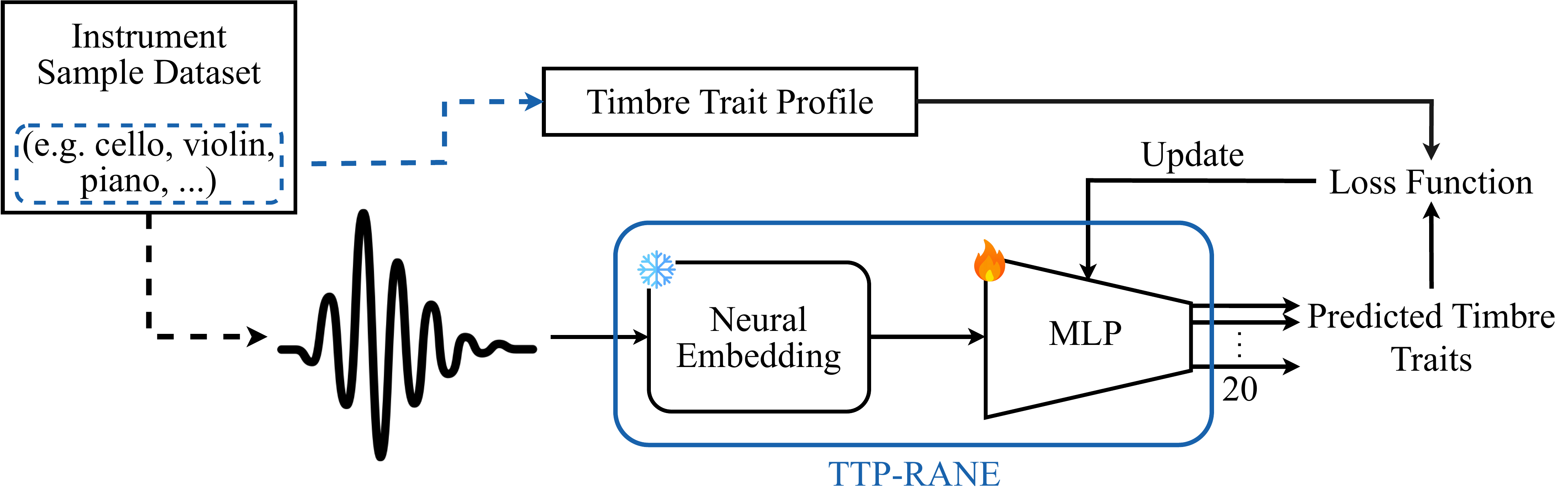}
        \caption{Timbre Traits Profiles - Reweighted Audio Neural Embedding (TTP-RANE): First, the frozen neural embedding model computes each audio sample embedding. Second, the trained MLP assigns values for the 20 timbre traits. During training, the loss function between predictions and ground truth (Timbre Trait Profiles) is computed to update the weights of the MLP. Solid lines indicates the process of the approach, dashed lines indicates where the dataset information is used.}
        \label{fig:Training Pipeline}
    \end{figure*}
    
    The evaluation of synthesized music instruments has become increasingly important as digital audio technology continues to advance. Whether for virtual instruments or AI-generated sounds, the ability to objectively measure how closely a synthesized sample matches its real-world counterpart is crucial for both developers and musicians. The standard for automatic assessment is the Fréchet Audio Distance (FAD) \cite{kilgour2019frechetaudiodistancemetric}. Originally inspired by the Fréchet Inception Distance used in image generation, FAD adapts this concept to the audio domain by comparing the statistical distributions of real and synthesized audio features. It provides a quantitative way to assess the perceptual quality and authenticity of synthesized instrument samples. To produce a meaningful metric, it requires embeddings that are trained on specific audio datasets—such as music or environmental sounds \cite{tailleur2024correlation, gui2024adapting}.

    Although FAD offers a valuable means of ranking the overall quality of synthesized samples, it remains a global metric with limited interpretability. Specifically, it does not provide actionable insights into which perceptual attributes might be responsible for discrepancies between real and synthesized sounds. This limitation underscores the need for more detailed and interpretable metrics, to our knowledge absent in the state-of-the-art, especially those related to timbre, as it is an essential dimension of audio perception that is difficult to formalize.

    Timbre has been the subject of extensive research in audio signal processing and music information retrieval. Because it requires bridging the gap between human perception of timbre and machine listening, timbre qualification remains challenging. Starting with pioneering work by Plomp (1970)\cite{plomp1970timbre}, Wessel (1973) \cite{wessel1973psychoacoustics}, and Grey (1977) \cite{grey1977multidimensional}, multidimensional scaling (MDS)—which uses pairwise dissimilarity ratings to model perceptual relationships among sounds—has become a common approach to mapping timbre space. MDS does not offer explanations of the resulting dimensions, which must instead be interpreted by the researchers. These dimensions are typically interpreted in terms of acoustic features; to the best of our knowledge, McAdams et al. (1995) \cite{mcadams1995perceptual} were the first to demonstrate that certain perceptual descriptors are correlated with the dimensions of the timbre space introduced in their study. Peeters et al. \cite{peeters2000instrument} proposed descriptors based on signal processing (e.g., spectral centroid, attack time) integrated into the MPEG-7 \cite{peeters2000instrument} audio description scheme for standardized timbre representation. Later, the Audio Commons initiative advanced the field by developing perceptual models linking signal processing metrics (e.g., zero crossing rate, spectral centroid, energy ratios) to high-level perceptual attributes (e.g., roughness, brightness, sharpness, warmth) \cite{pearce2019release,pearce2016first,pearce2017second}, offering a more interpretable timbre description, particularly for non-musical sounds. 
    The Timbre Toolbox \cite{peeters:hal-01106771, peeters_kazazis_mcadams_2023} emerged as a comprehensive resource, employing advanced techniques such as the Short-Time Fourier Transform and Constant-Q Transform to extract over 100 descriptors—including spectral, temporal, harmonic, and modulation features. Some of these acoustic features, like spectral centroid (linked to perceived brightness) and attack time (distinguishing percussive from sustained sounds), correlate with perceptual attributes \cite{schubert2006timbral,caetano2019audio}; however, relationships between acoustic correlates and semantic labels are often complex, with some research demonstrating significant advantages for nonlinear approaches \cite{mcadams2017perception, reymore2022modeling}. 
    
    Deep learning approaches, such as those by Pons et al. \cite{pons2017timbre}, introduced Convolutional Neural Networks (CNNs) to learn timbre representations directly from log-mel spectrograms, achieving state-of-the-art performance in tasks like instrument recognition and music auto-tagging. However, these features lack interpretability.

    In this paper, we address the lack of interpretable metrics for sound synthesizer assessment by introducing a predictor based on a 20-dimensional model of ``musical instrument timbre qualia" proposed by Reymore and Huron \cite{reymore2020timbre}. 
    
    Specifically, we leverage the data collected in this study to train a model to predict values across the 20 dimensions (called timbre traits) from audio input. This multi-layer-perceptron model is trained on deep neural embeddings of instrument samples from the RWC dataset, comprising 31 of the 34 instruments rated in a 2021 study by Reymore \cite{Reymore2021characterizing}. 
    
    We then demonstrate the benefit of this predictor for evaluating the performance of TokenSynth \cite{kim2025tokensynthtokenbasedneuralsynthesizer}, a sound synthesizer. First, we synthesize three sets of samples using three different conditionings of the model. We compute the predictions' Mean Absolute Error (MAE) on each of the synthesized sample set and demonstrate that it provides the same ranking as a FAD computed with the RWC database as a reference, suggesting that the proposed predictors are able to provide equivalent information on a distributional basis. Second, because the model is able to analyze sounds one by one, we can use it to analyze the timbral dimensions of generated sounds to improve the generation.

    The contributions of this paper are thus two-fold: 1) we introduce automatic timbre trait prediction as a machine listening task and demonstrate its feasibility; 2) we demonstrate the potential of our TTP prediction method ``Reweighted Audio Neural Embedding" (TTP-RANE) to complement FAD in order to obtain qualitative guidance for the evaluation of generative instrument audio synthesis.

    The paper is organized as follows. Section \ref{sec:timbre traits} introduces the work of Reymore and what timbre traits are. Section \ref{sec:predicting timbre traits} presents the training procedure and model performance. Section \ref{sec:synth assessment} presents the application of the model to evaluating synthesized sounds. Code and audio examples are made available.\footnote{Companion page: \url{https://theochaslecauchy.github.io/paperTTPSynthesizerAssessment/}}
    
    \begin{table}[t]
    \centering
    \begin{tabular}{|l|l|}
        \hline
        1. sparkling/brilliant  & 11. hollow \\
        2. focused/compact      & 12. woody \\ 
        3. pure/clear           & 13. airy/breathy \\
        4. open                 & 14. nasal/reedy \\
        5. ringing/long decay   & 15. raspy/grainy \\
        6. resonant/vibrant     & 16. rumbling/low \\
        7. sustained/even       & 17. direct/loud \\
        8. soft/singing         & 18. percussive \\
        9. watery/fluid         & 19. shrill/noisy \\
        10. muted/veiled        & 20. brassy/metallic \\ \hline
    \end{tabular}
    \caption{Dimensions of the Timbre Trait Profiles (TTP).}
    \label{tab:traits}
    \end{table}
    
    \section{Predicting timbre traits}\label{sec:predicting timbre traits}
    
    \subsection{Definition}\label{sec:timbre traits}

    The 20 dimensions of the timbre qualia model proposed by Reymore and Huron\cite{reymore2020timbre} were empirically derived through a two-stage process. First, open-ended interviews with 23 musicians generated 77 descriptive categories of timbre qualia, which were refined via content analysis and pile sorting. Second, 460 musicians rated these categories for 20 instruments. Principal Component Analysis (PCA) was applied to the ratings to reduce dimensionality, and feedback from musicians on component relevance was integrated, resulting in the 20 dimensions of the model. Reymore and Huron refer to the dimensions as timbre traits; these traits are presented in Table \ref{tab:traits}.

    
    In a subsequent study, Reymore \cite{Reymore2021characterizing} collected rating data from 243 musicians on each of the 20 dimensions in order to generate Timbre Trait Profiles (TTPs) for 34 common orchestral and wind ensemble instruments. As with the interviews and rating tasks from Reymore and Huron \cite{reymore2020timbre}, ratings used to create the TTPs were based on participants' imagined archetypical instrument sounds. For each instrument rated, participants were first asked to auralize\footnote{``imagine the sound of"} a typical sound played on that instrument, then to rate the imagined sound in the range $\llbracket1,7\rrbracket$ on each dimension of the timbre model. Each TTP includes the average participant ratings on 20 dimensions.
	
    The TTPs presented by Reymore in \cite{Reymore2021characterizing} were generated by averaging human ratings for each instrument. In this study, we present a TTP prediction method called TTP-RANE illustrated in Figure \ref{fig:Training Pipeline}. We trained a multi-layer perceptron model to predict TTPs from deep neural embeddings of instrument samples. To this end, we used the RWC audio dataset, with the TTPs assessed by Reymore \cite{Reymore2021characterizing} assigned to each sample according to its instrument class.
    
    \subsection{Dataset}
    
\begin{figure}[h!]
        \centering
        \includegraphics[width=\columnwidth]{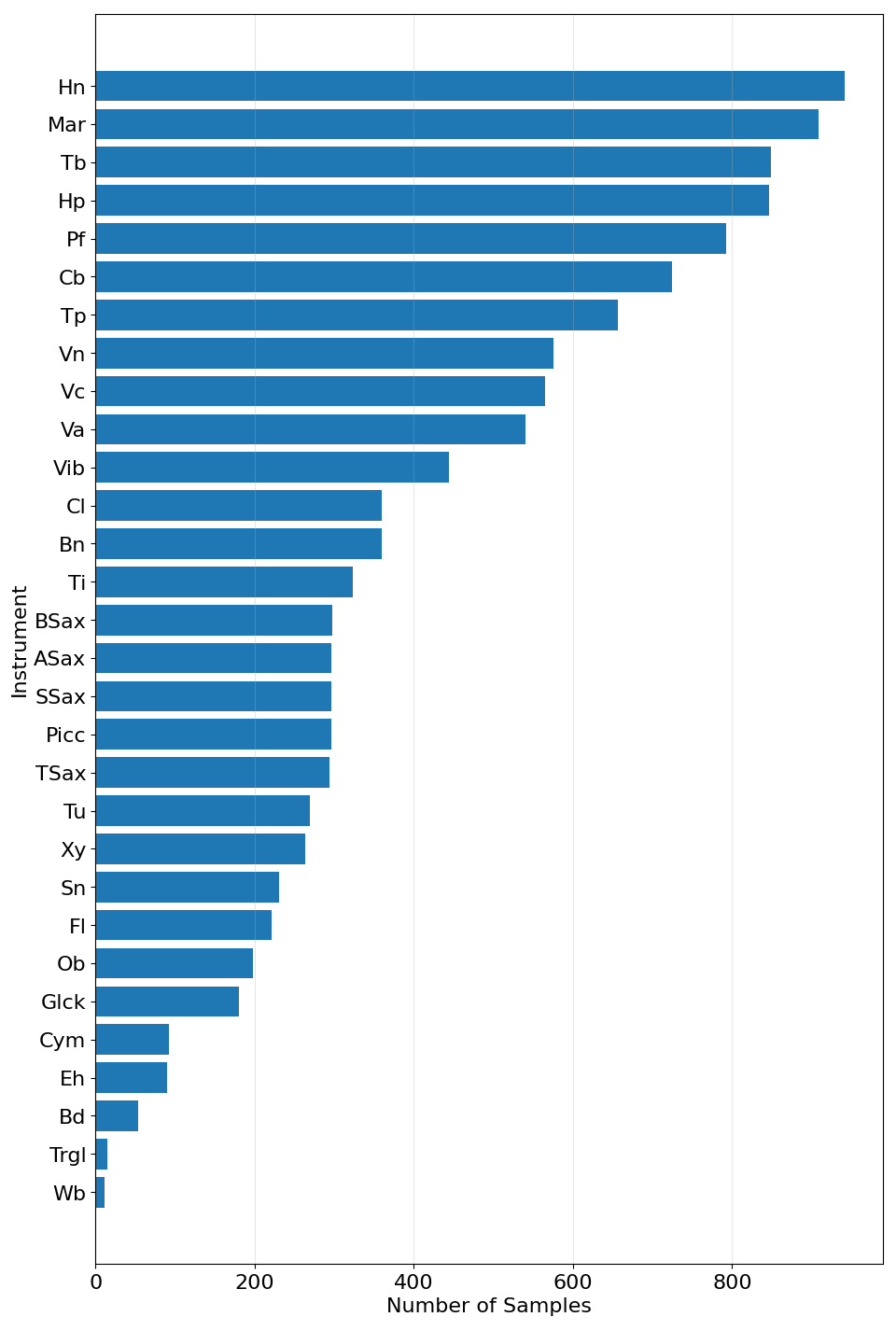}
        \caption{Musical instruments and their number of samples from RWC dataset.}
        \label{fig:instruments}
    \end{figure}

    The RWC dataset \cite{goto2003rwc} contains 91.6 hours of instrument samples, including 31 of the 34 musical instruments rated by Reymore \cite{Reymore2021characterizing}, listed in Figure \ref{fig:instruments}\footnote{Abbreviations correspondences can be found on the companion page.}. 
    
    For pitched instruments, each recording contains every note in the instrument’s pitch range, played successively. For percussive instruments (bass drum, cymbal, wood block, snare drum), which lack a defined pitch range, the recordings consist of several strikes or variations in playing style. Most instruments have recordings from different instrument brands or sub-instrument categories (e.g., for example for the piano, Pianoforte Yamaha, Steinway and Bosendorfer can be found). To obtain isolated note samples, each recording is segmented into individual notes. We first normalize samples and detected single notes with an initial threshold of -70 dB, decreasing it by 5dB successively if one of the resulting split audio files lasts more than 15 seconds. Only samples in their archetypical playing styles (ordinario) are considered for this study, in order to avoid playing styles that would diverge from the typical TTP of an instrument. The final number of extracted samples for each instrument is shown in Table \ref{fig:instruments}, corresponding to a total of 11 hours of audio. The average duration of the samples is 3.22 seconds and the standard deviations 1.4 seconds.

    \subsection{Neural Embeddings}
 
    We consider and compare predictors built upon 4 deep neural embeddings: VGGish \cite{simonyan2015deepconvolutionalnetworkslargescale}, the Music undERstanding model with large-scale self-supervised Training (MERT) \cite{li2024mertacousticmusicunderstanding}, Laion Contrastive Language-Audio Pretraining (CLAP) \cite{wu2024largescalecontrastivelanguageaudiopretraining} and its music-only version (CLAP-Music).

    VGGish \cite{simonyan2015deepconvolutionalnetworkslargescale} is an audio classifier trained on a subset of a large audio dataset extracted from YouTube videos, called YouTube-100M, which contains 350,000h of audio data with video-level class labels. VGGish adopts a CNN-based approach to extract features from log-mel spectrograms with 64 frequency bins and 10-ms hops to compute audio embeddings.
    
    MERT \cite{li2024mertacousticmusicunderstanding} is an embedding specialized in music analysis through a self-supervised learning framework using teacher-student methods: a combination of teachers, including an acoustic teacher based on Residual Vector Quantization - Variational AutoEncoder, a musical teacher based on the Constant-Q Transform, and a BERT-style transformer encoder as student model. The MERT models were trained on an extensive in-house private music dataset comprising 160,000 hours of audio data. Among these, the ``MERT-v1-95M" model stands out due to its performance on music understanding benchmarks.

    CLAP is trained to learn multi-modal representations using both an audio and a text encoder. First introduced in \cite{elizalde2022claplearningaudioconcepts} by Microsoft, LAION later presented a CLAP model with significantly better performance by training it on the large-scale LAION-Audio-630k and AudioSet datasets, enhanced with keyword-to-caption augmentation. Their best general-purpose model, ``630k-audioset-best", is trained on approximately 10,000 hours of audio and achieves state-of-the-art results on standard audio-text retrieval benchmarks, making it the most effective model for broad audio understanding.
    
    We also considered LAION CLAP models fine-tuned on music and speech data. Among these, the ``music audioset epoch 15 esc 90.14" model achieved superior performance on music-specific benchmark tasks. We hereafter refer to this model as ``CLAP-Music".

    We compute the CLAP, CLAP-Music, Mert and VGGish embeddings using the ``fadtk" python package considered by Gui et al. \cite{gui2024adaptingfrechetaudiodistance}.

    \subsection{Supervised approach}

    We present here a TTP prediction supervised approach called TTP-RANE. This method, illustrated in Figure \ref{fig:Training Pipeline}, is composed of a frozen deep neural embedding following by a shallow learnable MLP reweighting the neural embedding. "Reweighting" refers here to the process of using the MLP to recalibrate the weights of the neural embedding features to achieve this prediction task. We train this MLP to predict TTPs from audio embeddings of RWC dataset samples.
    
    To this end, the embeddings dataset is shuffled and split into a training and a validation set (80\%-20\%). We use the TTP as ground truth labels. The values of timbre traits \cite{Reymore2021characterizing} belong to the range $\llbracket1, 7\rrbracket$. First, the values are normalized to the interval [0,1], and a sigmoid is applied to the output values of the model to predict values in this interval. We used the Mean Square Error (MSE), commonly applied in regression problems, as our loss function. 
    
    Given the limited amount of ground truth data, we evaluate the models' performance using a cross-evaluation approach based on instrument. That is, for each instrument, we train a model using samples from all the other instruments and then evaluate its performance on samples from the target instrument. To evaluate a given model, we thus train 31 separate models, one for each instrument, with the same architecture and training conditions, and we evaluate each model on all its unknown instrument samples (training set and validation set).
    
    We test 3 MLP architectures: 1) one with no hidden layer, 2) one with one hidden layer of size 256, and 3) a funnel-shaped architecture with 2 hidden layers of sizes 256 and 128. We use a learning rate of 0.01 scheduled with a ``Reduce on Plateau" scheduler, an Adam optimizer and a maximum of 300 epochs per training. We apply early stopping on the validation set with a patience of 20 epochs; that is, the training stops when the loss of the predictions on the validation set does not decrease during 20 consecutive epochs, and we keep the model with the lowest validation loss.

    \subsection{Unsupervised approach} 
    
    To measure to benefit of using TTP-RANE, we consider an unsupervised approach that does not consider any specialized prior knowledge on musical timbre, called ``Text-to-Audio Similarity" (T2ASim). T2ASim predicts timbre trait values by leveraging the similarity between audio and text embeddings in the CLAP space. Specifically, for an audio sample with a CLAP embedding $\mathbf{E_i}$ and a text embedding $\mathbf{t}_j$ representing the $j$-th trait, the prediction process is as follows:

    \begin{enumerate}
        \item Embeddings Distance: Computation of the distance $d(\mathbf{E_i}, \mathbf{t}_j)$ between the audio embedding $\mathbf{E_i}$ and the trait's text embedding $\mathbf{t}_j$.
        \item Normalization: Divide the distance by $\max_{i,j}(d(\mathbf{E_i}, \mathbf{t}_j))$.
        \item Prediction: Derive the predicted trait value as $1 - d(\mathbf{E_i}, \mathbf{t}_j)$.
    \end{enumerate}

    \subsection{Results}    
    To assess the performance of the evaluated models, we consider the Pearson correlation between the predicted TTPs and the ground truth TTPs. Each element of the prediction array is the average of the predictions for a specific timbre trait and instrument in the RWC dataset. As there are 20 timbre traits and 31 instruments, this results in 620 points per model for the correlation computation. The higher the Pearson correlation values, the better the timbre trait values prediction.

    Cross-evaluation results are displayed in Table \ref{tab:performances_table}. For each embedding type, Table \ref{tab:performances_table} presents only the model with no hidden layer, as we found it to be the architecture with the best cross-evaluation performance. We observe on the training loss that the models with 1 or more hidden layers overfit on the training set resulting in lower performances during the cross-evaluation, contrary to the model with no hidden layer that generalizes better to instruments not included in the training set.

    \begin{table}[t]
        \centering
        \begin{tabular}{ll>{\raggedleft\arraybackslash}r}
            \toprule
             & Embedding & Pearson $\uparrow$\\
            \midrule
            T2ASim & CLAP \cite{wu2024largescalecontrastivelanguageaudiopretraining} & .101$^{*}$\\
            \midrule
            \multirow{4}{*}{TTP-RANE} & MERT \cite{li2024mertacousticmusicunderstanding} & .578$^{*}$\\
            & VGGish \cite{simonyan2015deepconvolutionalnetworkslargescale} & .581$^{*}$\\
             & CLAP-Mus. \cite{wu2024largescalecontrastivelanguageaudiopretraining} & .631$^{*}$\\
             & CLAP \cite{wu2024largescalecontrastivelanguageaudiopretraining} & \textbf{.663$^{*}$}\\
            \midrule
            Hum. Ratings & & .698$^{*}$\\
            \bottomrule
        \end{tabular}
        \caption{Comparison of Pearson correlations of the predicted timbre traits values by models and human ratings (* p value under 0.001).}
        \label{tab:performances_table}
    \end{table}
    
    For the human ratings, the Pearson correlation is computed between the ratings collected by Reymore \cite{Reymore2021characterizing} and the ground truth values (that correspond to the average human ratings). This gives an estimate of the inter-rater agreement of human ratings.\footnote{Inter-rater agreement refers to the degree of consistency or concordance among ratings provided by different human evaluators.}
    
    The model achieving the best cross-evaluation performance is the model trained on CLAP embeddings. Thus for the assessment of the synthesizer, we train a new model on all 31 instruments using CLAP embeddings and the same parameters as previously, referring to it hereafter as TTP-RANE-CLAP.

	\section{Assessing TokenSynth}\label{sec:synth assessment}
    In this section, we demonstrate the potential of using the timbre trait predictor to assess the performance of a musical instrument synthesizer: TokenSynth.
    
    \subsection{Overall Synthesis Assessment}
    TokenSynth \cite{kim2025tokensynthtokenbasedneuralsynthesizer} is a neural synthesizer that leverages token-based audio representations to enable text-to-instrument synthesis. This synthesizer takes as input a MIDI token and a CLAP embedding. It employs a decoder-only transformer that computes audio tokens from inputs, decoded by a pre-trained Descript Audio Codec (DAC) to generate audio samples. The model was trained on a dataset combining NSynth \cite{engel2017neuralaudiosynthesismusical} and Lakh MIDI \cite{raffel2016learning}, totaling over 9.53 million samples, which was further augmented with digital audio effects to enhance timbral diversity. 
    
    For each instrument in Table \ref{fig:instruments}, we randomly sample 100 pitches in the corresponding pitch range of the instrument recordings in the RWC dataset (e.g., G3 to E7 for the violin). For each instrument, we sample pitches from a Gaussian distribution whose mean is set at the midpoint of the instrument’s pitch range and whose standard deviation is set to 20\% of the total pitch range. CLAP embeddings enable the projection of text and audio in the same space. As TokenSynth synthesis is conditioned with a CLAP embedding, we can use text and/or audio conditioning. Therefore, for each sampled pitch of an instrument $\mathbf{I}$, we synthesize 3 sounds with different embeddings as input: 1) Text conditioning: the text embedding of the instrument $\mathbf{I}$ name (e.g., ``CELLO", ``SNARE DRUM"); 2) Audio conditioning: the averaged audio embedding of all samples of the instrument $\mathbf{I}$  from the RWC dataset; 3) Text-Audio conditioning: the mean embedding of both previous embeddings.
    
    To assess the synthesizer, we first compute FAD between RWC samples and TokenSynth samples, then we compute the MAE between predicted TTPs on synthesized samples and the ground truth. For the sake of comparison, we also compute the MAE between predicted TTPs on the RWC samples (training set and validation set) and the ground truth.
    
    Results are shown in Table \ref{tab:synth_assessment}. We observe that there are no significant statistical difference between different conditioning generations with respect to their MAE.  Furthermore, MAE between predicted TTPs and the ground truth gives the same ranking as the FAD, suggesting that TTP-RANE-CLAP provides, at a macro scale, equivalent information about the performance of the studied synthesizer. We can now go deeper in the analysis. In the following, we focus on the audio conditioned synthesis.

    \begin{table}
        \centering
        \begin{tabular}{lrr}
            \toprule
             & FAD $\downarrow$ & MAE $\downarrow$\\
            \midrule
            RWC samples &  & .058 ± .056\\
            \midrule
            Text conditioned samples & .57 & .180 ± .128\\
            Audio conditioned samples & .53 & \textbf{.172 ± .123}\\
            Text-Audio cond. samples & .54 & .173 ± .124\\
            \bottomrule
        \end{tabular}
        \caption{FAD between RWC and synthesized samples and MAE of the predicted timbre trait values on the synthesized samples compared to the MAE on the RWC dataset.}
        \label{tab:synth_assessment}
    \end{table}
    
    \subsection{Wood Block Synthesis Assessment}
    We can start the analysis of this synthesis by observing the predicted TTP of the instrument with the highest MAE, here, the wood block (MAE of 0.24). This indicates that the wood block is the instrument which synthesized samples have the furthest predicted TTPs from the ground truth TTP. The wood block is the less represented instrument in the RWC dataset, this could lead to less accurate predictions by TTP-RANE. However, the investigation on synthesized wood block samples confirms these results. Figure \ref{fig:Synth Wood Block TTP} shows the predicted TTP of the synthesized wood block, that is, the average predicted TTP across all samples. 

    \begin{figure}[h!]
        \centering
        \resizebox{\columnwidth}{!}{
\begin{tikzpicture}
  \begin{polaraxis}[
    yticklabel style={/pgf/number format/fixed},
    yticklabels={0.2, , .2, .4, .6, .8},
    xticklabels={\shortstack{sparkling\\brilliant}, \shortstack{focused\\compact}, \shortstack{pure\\clear}, open, \shortstack{ringing\\long decay}, \shortstack{resonant\\vibrant}, \shortstack{sustained\\even}, \shortstack{soft\\singing}, \shortstack{watery\\fluid}, \shortstack{muted\\veiled}, hollow, woody, \shortstack{airy\\breathy}, \shortstack{nasal\\reedy}, \shortstack{raspy\\grainy}, \shortstack{rumbling\\low}, \shortstack{direct\\loud}, percussive, \shortstack{shrill\\noisy}, \shortstack{brassy\\metallic}},
    xtick={0.0, 18.0, 36.0, 54.0, 72.0, 90.0, 108.0, 126.0, 144.0, 162.0, 180.0, 198.0, 216.0, 234.0, 252.0, 270.0, 288.0, 306.0, 324.0, 342.0},
    xticklabel style={
      inner sep=5pt,
      font=\small,
    },
    grid=both,
    axis line style={draw=none}, 
    ymin=0,
    ymax=1,
    legend pos=outer north east,
    legend style={at={(0.5, -0.3)}, anchor=south},
  ]
    \addplot[black, thick, domain=0:360, samples=100] {1};
    
    \addplot[green!70!black, opacity=0.25, name path=upper_gv] coordinates {
          (0.0, 0.2308005239981721)
      (18.0, 0.8622750046491958)
      (36.0, 0.7718010514282247)
      (54.0, 0.4949331624263167)
      (72.0, 0.1812226856611664)
      (90.0, 0.37780707740402725)
      (108.0, 0.16441824536637661)
      (126.0, 0.07076181230660968)
      (144.0, 0.06521618008754297)
      (162.0, 0.22653064093781056)
      (180.0, 0.7818380884722974)
      (198.0, 0.9156008624846004)
      (216.0, 0.13864094011500888)
      (234.0, 0.19185005864544077)
      (252.0, 0.0765326398657555)
      (270.0, 0.1008706393046486)
      (288.0, 0.7215971366351666)
      (306.0, 0.9868951561941233)
      (324.0, 0.5051609409171476)
      (342.0, 0.033230288517987344)
      (0.0, 0.2308005239981721)
    };

    \addplot[green!70!black, opacity=0.25, name path=lower_gv] coordinates {
          (0.0, 0.11253280933516116)
      (18.0, 0.7343916620174709)
      (36.0, 0.604865615238442)
      (54.0, 0.3417335042403499)
      (72.0, 0.08211064767216697)
      (90.0, 0.22219292259597273)
      (108.0, 0.05891508796695669)
      (126.0, 0.01590485436005699)
      (144.0, 0.01145048657912369)
      (162.0, 0.10013602572885612)
      (180.0, 0.6348285781943693)
      (198.0, 0.787732470848733)
      (216.0, 0.041359059884991126)
      (234.0, 0.07148327468789259)
      (252.0, 0.010134026800911175)
      (270.0, 0.02579602736201802)
      (288.0, 0.5584028633648335)
      (306.0, 0.9131048438058769)
      (324.0, 0.32150572574951913)
      (342.0, 0.00010304481534601645)
      (0.0, 0.11253280933516116)
    };

    \addplot[green!10] fill between[of=upper_gv and lower_gv];

    \node[anchor=south west] at (axis cs: 177,0.7) {\textcolor{green!40!black}{G}};
    \addplot[green!70!black, opacity=0.75] coordinates {
          (0.0, 0.17166666666666663)
      (18.0, 0.7983333333333333)
      (36.0, 0.6883333333333334)
      (54.0, 0.4183333333333333)
      (72.0, 0.13166666666666668)
      (90.0, 0.3)
      (108.0, 0.11166666666666665)
      (126.0, 0.043333333333333335)
      (144.0, 0.03833333333333333)
      (162.0, 0.16333333333333333)
      (180.0, 0.7083333333333334)
      (198.0, 0.8516666666666667)
      (216.0, 0.09000000000000001)
      (234.0, 0.13166666666666668)
      (252.0, 0.043333333333333335)
      (270.0, 0.06333333333333331)
      (288.0, 0.64)
      (306.0, 0.9500000000000001)
      (324.0, 0.41333333333333333)
      (342.0, 0.01666666666666668)
      (0.0, 0.17166666666666663)
    };

    \addplot[blue!70!black, opacity=0.25, name path=upper] coordinates {
          (0.0, 0.3817182270530651)
      (18.0, 0.4986495780397354)
      (36.0, 0.5507650254542966)
      (54.0, 0.5381465850788061)
      (72.0, 0.4685872038356402)
      (90.0, 0.6688823560860399)
      (108.0, 0.5396209808324445)
      (126.0, 0.43905082262625056)
      (144.0, 0.3298260634241623)
      (162.0, 0.3531214217510912)
      (180.0, 0.40331681860491625)
      (198.0, 0.42911634303931695)
      (216.0, 0.3373704697867699)
      (234.0, 0.41977520408850283)
      (252.0, 0.39039143907633067)
      (270.0, 0.5105866880424547)
      (288.0, 0.5144523274955594)
      (306.0, 0.4953619895697333)
      (324.0, 0.28924977331977897)
      (342.0, 0.3328976966060165)
      (0.0, 0.3817182270530651)
    };

    \addplot[blue!70!black, opacity=0.25, name path=lower] coordinates {
          (0.0, 0.3693290313240101)
      (18.0, 0.49047564273586874)
      (36.0, 0.5412190319721562)
      (54.0, 0.5299150295538004)
      (72.0, 0.4547726265438459)
      (90.0, 0.6614548488471267)
      (108.0, 0.5284548256660832)
      (126.0, 0.42095356069932616)
      (144.0, 0.32066534766573473)
      (162.0, 0.3447154666814116)
      (180.0, 0.3950652848382295)
      (198.0, 0.4061455949365666)
      (216.0, 0.3267988161066704)
      (234.0, 0.3985996973254147)
      (252.0, 0.37220737111959223)
      (270.0, 0.4924913119070006)
      (288.0, 0.5043426030817187)
      (306.0, 0.4764003290414117)
      (324.0, 0.276158072058143)
      (342.0, 0.31474536914458234)
      (0.0, 0.3693290313240101)
    };

    \addplot[blue!10] fill between[of=upper and lower];

    \node[anchor=south west] at (axis cs: 120,0.4) {\textcolor{blue!50!black}{B}};
    \addplot[blue!70!black, opacity=0.75] coordinates {
          (0.0, 0.3755236291885376)
      (18.0, 0.4945626103878021)
      (36.0, 0.5459920287132264)
      (54.0, 0.5340308073163033)
      (72.0, 0.46167991518974305)
      (90.0, 0.6651686024665833)
      (108.0, 0.5340379032492638)
      (126.0, 0.43000219166278836)
      (144.0, 0.3252457055449485)
      (162.0, 0.3489184442162514)
      (180.0, 0.3991910517215729)
      (198.0, 0.4176309689879418)
      (216.0, 0.33208464294672013)
      (234.0, 0.4091874507069588)
      (252.0, 0.38129940509796145)
      (270.0, 0.5015389999747276)
      (288.0, 0.509397465288639)
      (306.0, 0.4858811593055725)
      (324.0, 0.282703922688961)
      (342.0, 0.3238215328752994)
      (0.0, 0.3755236291885376)
    };

    \node[anchor=south west] at (axis cs: 120,0.6666666) {\textcolor{red!60!black}{R}};
    \addplot[red!70!black, opacity=0.75] coordinates {
          (0.0, 0.4299322664737701)
      (18.0, 0.4640472531318664)
      (36.0, 0.5461021065711975)
      (54.0, 0.5609493255615234)
      (72.0, 0.4998268485069275)
      (90.0, 0.7254638671875)
      (108.0, 0.6270599961280823)
      (126.0, 0.541641354560852)
      (144.0, 0.3890601396560669)
      (162.0, 0.3787133395671844)
      (180.0, 0.3981121182441711)
      (198.0, 0.4898581802845001)
      (216.0, 0.4255733489990234)
      (234.0, 0.5085607171058655)
      (252.0, 0.3677980899810791)
      (270.0, 0.4855550229549408)
      (288.0, 0.4790868759155273)
      (306.0, 0.3414754271507263)
      (324.0, 0.2794384956359863)
      (342.0, 0.3179948925971985)
      (0.0, 0.4299322664737701)
    };

  \end{polaraxis}
\end{tikzpicture}
}
        \vspace{-1cm}
        \caption{Radar plots of TTP for Wood Block with  their respective 95\% confidence interval for Reymore's results (Green); Averaged predicted TTP on synthesized wood block samples (Blue); Predicted TTP of the sample with highest difference between ``percussive" prediction and ground truth (Red).}
        \label{fig:Synth Wood Block TTP}
    \end{figure}
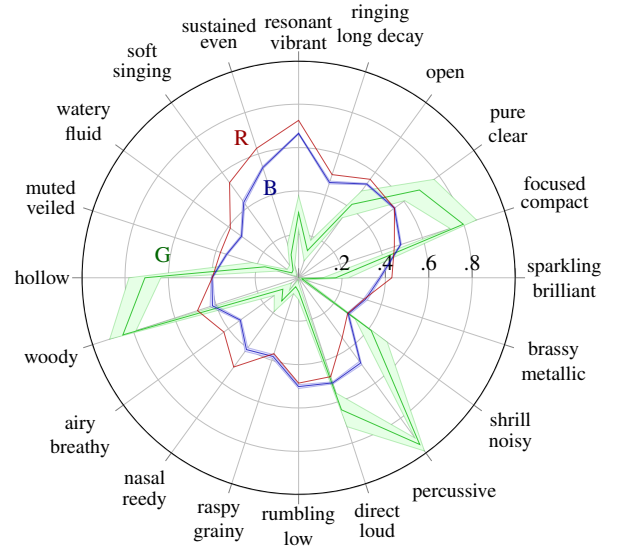
    \begin{figure}[h!]
        \centering
        \subfloat{
            \includegraphics[width=\columnwidth]{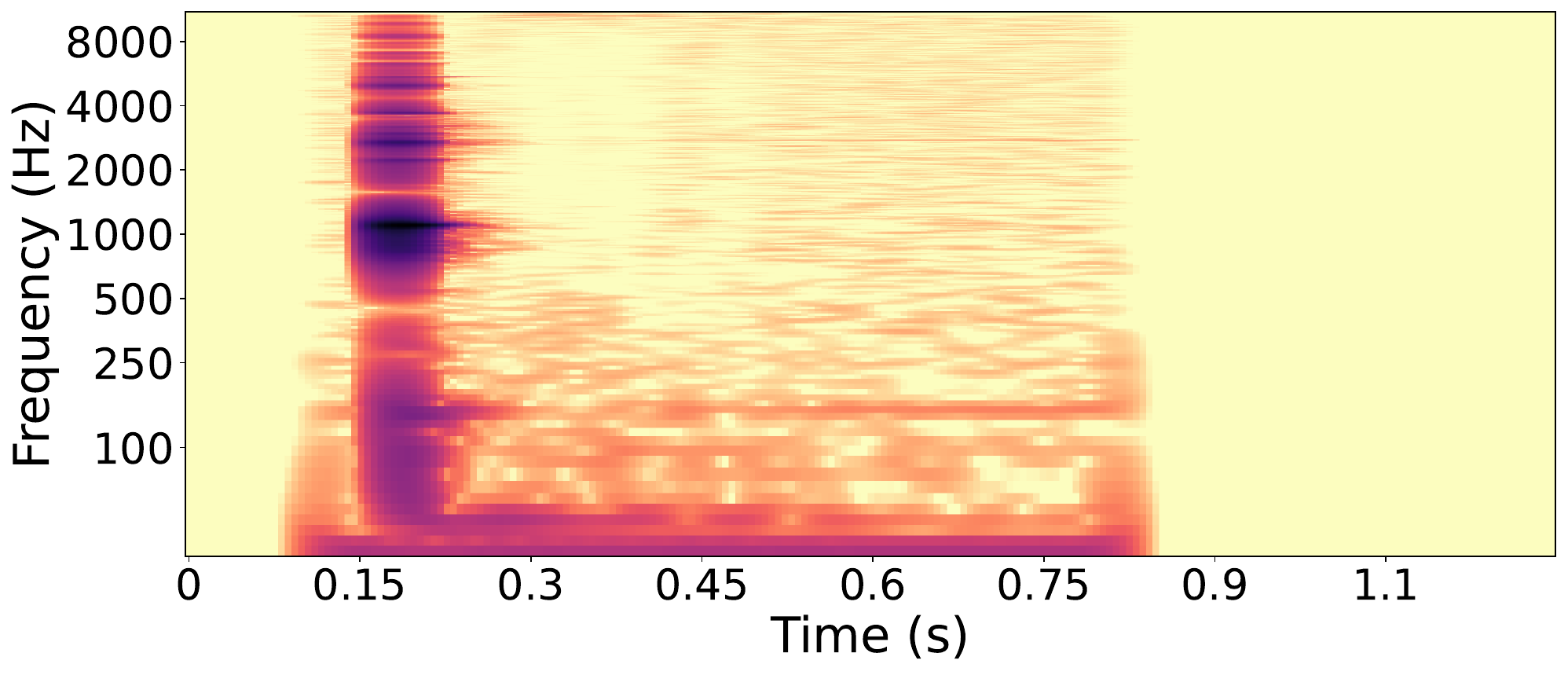}
        }
        \hfill
        \subfloat{
            \includegraphics[width=\columnwidth]{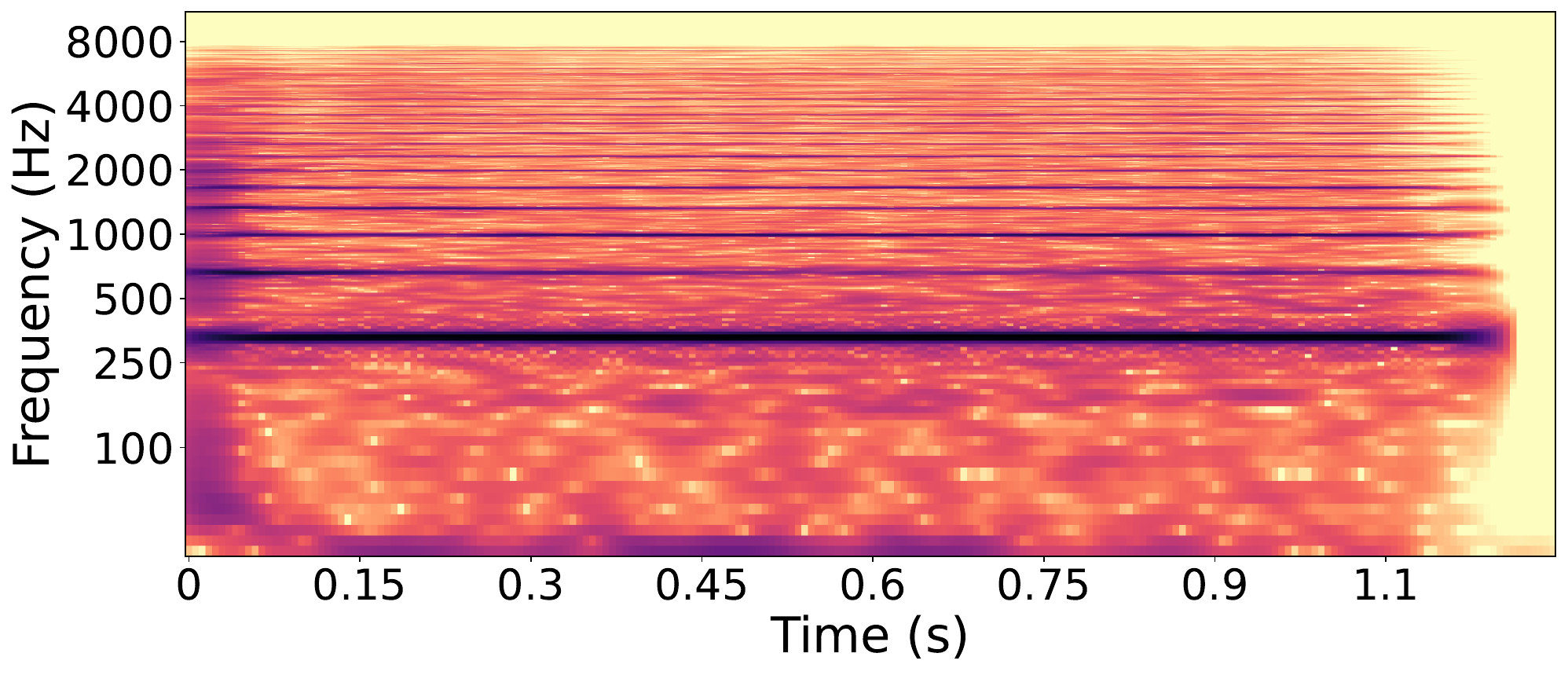}
        }
        \caption{Above: Spectrogram of the RWC wood block sample with the lowest difference between predicted ``percussive" value and ground truth. Below: Spectrogram of the synthesized wood block sample with the highest difference between predicted ``percussive" value and ground truth}
        \label{fig:Synth Wood Block Spectrograms Percussive}
    \end{figure}

    We observe on the wood block averaged predicted TTP that the timbre traits with the highest errors are ``woody" and ``percussive." While ``woody" can not be easily assessed on a spectrogram by a non-specialist, we can make observations with respect to the ``percussive" trait. Figure \ref{fig:Synth Wood Block Spectrograms Percussive} shows the spectrograms of a) the synthesized wood block sample with the highest difference between predicted ``percussive" value and ground truth, and b) the RWC wood block sample with the lowest difference between predicted ``percussive" value and ground truth. 
    
    The wood block is a percussive instrument, so its ground truth value of the ``percussive" trait is about 0.93. The prediction of this trait for the synthesized sample is 0.32. We can observe on the spectrograms that the synthesized sample is sustained, contrary to the RWC sample, which decays quickly after impact. This explains why its ``percussive" trait prediction is far below the ground truth.

    \subsection{Cello Synthesis Assessment}

    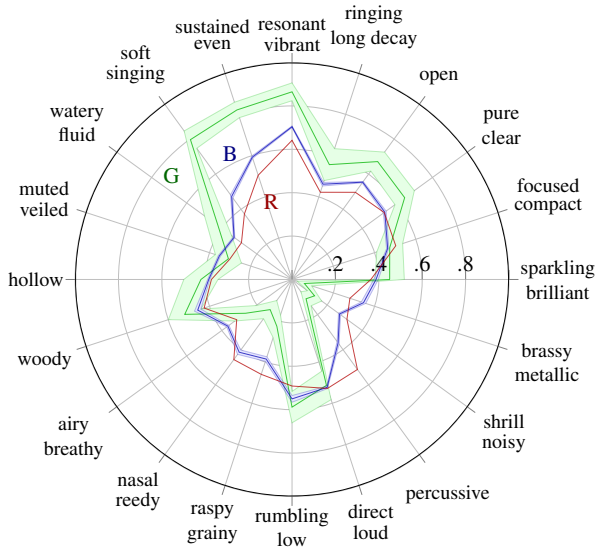
\begin{figure}[h!]
        \centering
        \resizebox{\columnwidth}{!}{
\begin{tikzpicture}
  \begin{polaraxis}[
    yticklabel style={/pgf/number format/fixed},
    yticklabels={0.2, , .2, .4, .6, .8},
    xticklabels={\shortstack{sparkling\\brilliant}, \shortstack{focused\\compact}, \shortstack{pure\\clear}, open, \shortstack{ringing\\long decay}, \shortstack{resonant\\vibrant}, \shortstack{sustained\\even}, \shortstack{soft\\singing}, \shortstack{watery\\fluid}, \shortstack{muted\\veiled}, hollow, woody, \shortstack{airy\\breathy}, \shortstack{nasal\\reedy}, \shortstack{raspy\\grainy}, \shortstack{rumbling\\low}, \shortstack{direct\\loud}, percussive, \shortstack{shrill\\noisy}, \shortstack{brassy\\metallic}},
    xtick={0.0, 18.0, 36.0, 54.0, 72.0, 90.0, 108.0, 126.0, 144.0, 162.0, 180.0, 198.0, 216.0, 234.0, 252.0, 270.0, 288.0, 306.0, 324.0, 342.0},
    xticklabel style={
      inner sep=5pt,
      font=\small,
    },
    grid=both,
    axis line style={draw=none},
    ymin=0,
    ymax=1,
    legend pos=outer north east,
    legend style={at={(0.5, -0.3)}, anchor=south},
  ]
    \addplot[black, thick, domain=0:360, samples=100] {1};
    \addplot[green!70!black, opacity=0.25, name path=upper_gv] coordinates {
          (0.0, 0.5176433430533076)
      (18.0, 0.536097052669619)
      (36.0, 0.6967307991539012)
      (54.0, 0.7257266297379962)
      (72.0, 0.6339835492412765)
      (90.0, 0.9061281259220183)
      (108.0, 0.8590021040808772)
      (126.0, 0.8461271462373136)
      (144.0, 0.49868440810080406)
      (162.0, 0.37130939176844097)
      (180.0, 0.498848481868556)
      (198.0, 0.5993197271582427)
      (216.0, 0.3257323693296965)
      (234.0, 0.22342424517942014)
      (252.0, 0.2816195685826267)
      (270.0, 0.6607311228019004)
      (288.0, 0.5824812203760686)
      (306.0, 0.14990629277698114)
      (324.0, 0.16889630790888707)
      (342.0, 0.09511668864430346)
      (0.0, 0.5176433430533076)
    };

    \addplot[green!70!black, opacity=0.25, name path=lower_gv] coordinates {
      (0.0, 0.38235665694669246)
      (18.0, 0.4105696139970476)
      (36.0, 0.586602534179432)
      (54.0, 0.6176067035953373)
      (72.0, 0.4793497840920567)
      (90.0, 0.8238718740779819)
      (108.0, 0.7843312292524561)
      (126.0, 0.7472061870960199)
      (144.0, 0.3446489252325292)
      (162.0, 0.24535727489822573)
      (180.0, 0.34115151813144395)
      (198.0, 0.4406802728417574)
      (216.0, 0.2042676306703034)
      (234.0, 0.11990908815391313)
      (252.0, 0.1717137647507066)
      (270.0, 0.5159355438647664)
      (288.0, 0.444185446290598)
      (306.0, 0.07009370722301883)
      (324.0, 0.08777035875777958)
      (342.0, 0.024883311355696575)
      (0.0, 0.38235665694669246)
    };

    \addplot[green!10] fill between[of=upper_gv and lower_gv, reverse=true];
    
    \addplot[green!70!black, opacity=0.75] coordinates {
          (0.0, 0.45)
      (18.0, 0.47333333333333333)
      (36.0, 0.6416666666666666)
      (54.0, 0.6716666666666667)
      (72.0, 0.5566666666666666)
      (90.0, 0.8650000000000001)
      (108.0, 0.8216666666666667)
      (126.0, 0.7966666666666667)
      (144.0, 0.42166666666666663)
      (162.0, 0.30833333333333335)
      (180.0, 0.42)
      (198.0, 0.52)
      (216.0, 0.26499999999999996)
      (234.0, 0.17166666666666663)
      (252.0, 0.22666666666666666)
      (270.0, 0.5883333333333334)
      (288.0, 0.5133333333333333)
      (306.0, 0.10999999999999999)
      (324.0, 0.12833333333333333)
      (342.0, 0.06000000000000002)
      (0.0, 0.45)
    };

    \node[anchor=south west] at (axis cs: 147,0.75) {\textcolor{green!40!black}{G}};

    \addplot[blue!70!black, opacity=0.25, name path=upper] coordinates {
          (0.0, 0.3977127140575498)
      (18.0, 0.46896065781432544)
      (36.0, 0.5362570787918077)
      (54.0, 0.5582667117449661)
      (72.0, 0.4704697752669525)
      (90.0, 0.7078509569801779)
      (108.0, 0.598138251249951)
      (126.0, 0.48263342757628863)
      (144.0, 0.3345015447734309)
      (162.0, 0.35727598264369725)
      (180.0, 0.38908559501399403)
      (198.0, 0.47261961197052754)
      (216.0, 0.3717653063817318)
      (234.0, 0.4268199729028522)
      (252.0, 0.39667449890261436)
      (270.0, 0.5657087340428841)
      (288.0, 0.5287532512783889)
      (306.0, 0.36890931489853196)
      (324.0, 0.27631728699258895)
      (342.0, 0.3571168195485131)
      (0.0, 0.3977127140575498)
    };

    \addplot[blue!70!black, opacity=0.25, name path=lower] coordinates {
          (0.0, 0.376517766136542)
      (18.0, 0.4579983352629527)
      (36.0, 0.5205971483934417)
      (54.0, 0.5504386354354004)
      (72.0, 0.4531689905449677)
      (90.0, 0.7003810405097515)
      (108.0, 0.587219426686603)
      (126.0, 0.4642014823396354)
      (144.0, 0.32589675790086253)
      (162.0, 0.34709447488632444)
      (180.0, 0.37891934007177946)
      (198.0, 0.4417605842506452)
      (216.0, 0.3596788336949055)
      (234.0, 0.40208404415814286)
      (252.0, 0.3746341653298781)
      (270.0, 0.5349906459972846)
      (288.0, 0.5171191455483551)
      (306.0, 0.35338630375476543)
      (324.0, 0.26458123961635494)
      (342.0, 0.33564484305391146)
      (0.0, 0.376517766136542)
    };

    \addplot[blue!10] fill between[of=upper and lower];

    \node[anchor=south west] at (axis cs: 125,0.62) {\textcolor{blue!50!black}{B}};
    \addplot[blue!70!black, opacity=0.75] coordinates {
          (0.0, 0.3871152400970459)
      (18.0, 0.46347949653863907)
      (36.0, 0.5284271135926247)
      (54.0, 0.5543526735901833)
      (72.0, 0.4618193829059601)
      (90.0, 0.7041159987449647)
      (108.0, 0.592678838968277)
      (126.0, 0.473417454957962)
      (144.0, 0.3301991513371467)
      (162.0, 0.35218522876501085)
      (180.0, 0.38400246754288675)
      (198.0, 0.45719009811058636)
      (216.0, 0.36572207003831864)
      (234.0, 0.4144520085304975)
      (252.0, 0.3856543321162462)
      (270.0, 0.5503496900200844)
      (288.0, 0.522936198413372)
      (306.0, 0.3611478093266487)
      (324.0, 0.27044926330447194)
      (342.0, 0.3463808313012123)
      (0.0, 0.3871152400970459)
    };

    \node[anchor=south west] at (axis cs: 120,0.333333) {\textcolor{red!60!black}{R}};
    \addplot[red!70!black, opacity=0.75] coordinates {
          (0.0, 0.3633850812911987)
      (18.0, 0.5027209520339966)
      (36.0, 0.5249559879302979)
      (54.0, 0.496666818857193)
      (72.0, 0.422670841217041)
      (90.0, 0.641447126865387)
      (108.0, 0.5050306916236877)
      (126.0, 0.3734101951122284)
      (144.0, 0.2882046997547149)
      (162.0, 0.3021149039268493)
      (180.0, 0.3715859055519104)
      (198.0, 0.426090806722641)
      (216.0, 0.3152448236942291)
      (234.0, 0.4564326703548431)
      (252.0, 0.459712415933609)
      (270.0, 0.4910845458507538)
      (288.0, 0.528049886226654)
      (306.0, 0.5127891898155212)
      (324.0, 0.3127066195011139)
      (342.0, 0.2804427444934845)
      (0.0, 0.3633850812911987)
    };

  \end{polaraxis}
\end{tikzpicture}
}
        \vspace{-1cm}
        \caption{Radar plots of TTP for Cello with their respective 95\% confidence interval for Reymore's results (Green); Averaged predicted TTP on synthesized cello samples (Blue); Predicted TTP of the sample with highest difference between ``resonant-vibrant" prediction and ground truth (Red).}
        \label{fig:Synth Cello TTP}
    \end{figure}

    \begin{figure}[h!]
        \centering
        \subfloat{
            \includegraphics[width=\columnwidth]{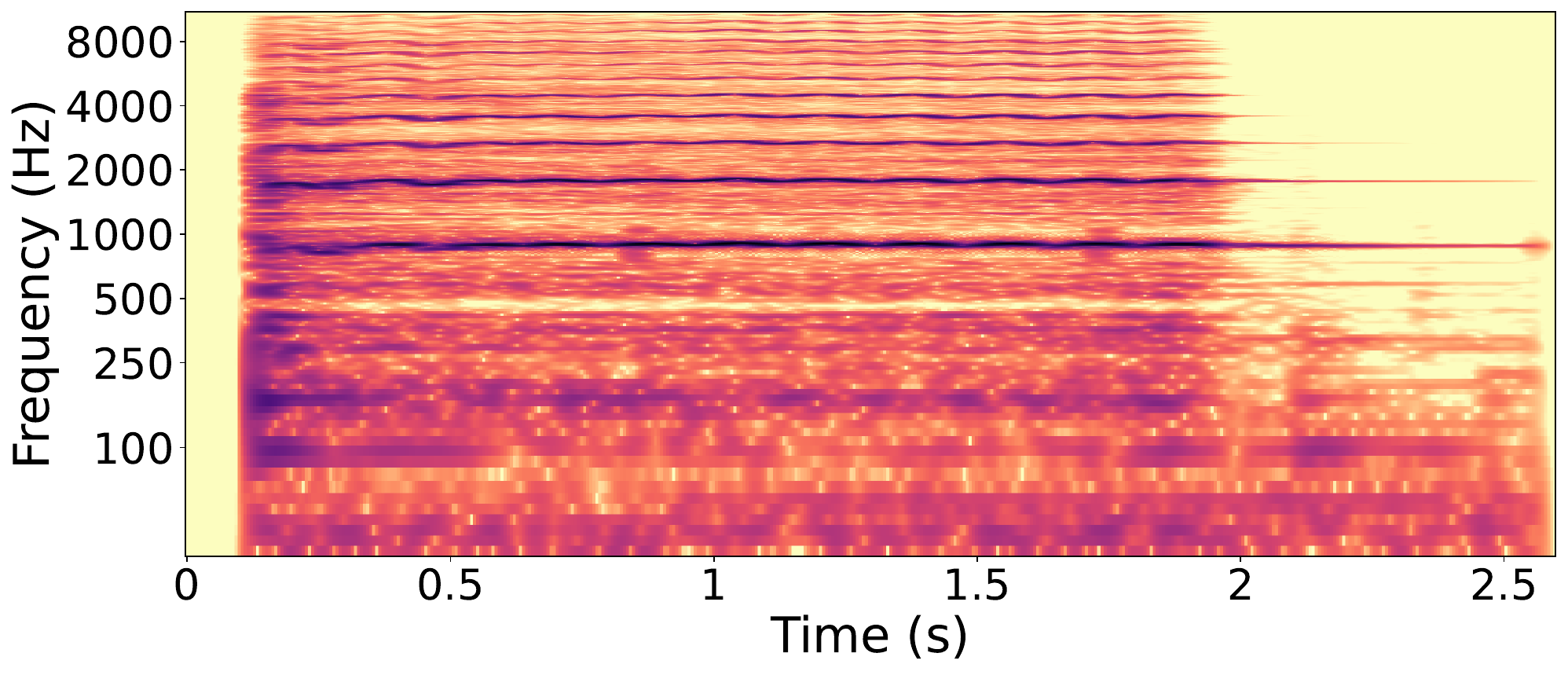}
        }
        \hfill
        \subfloat{
            \includegraphics[width=\columnwidth]{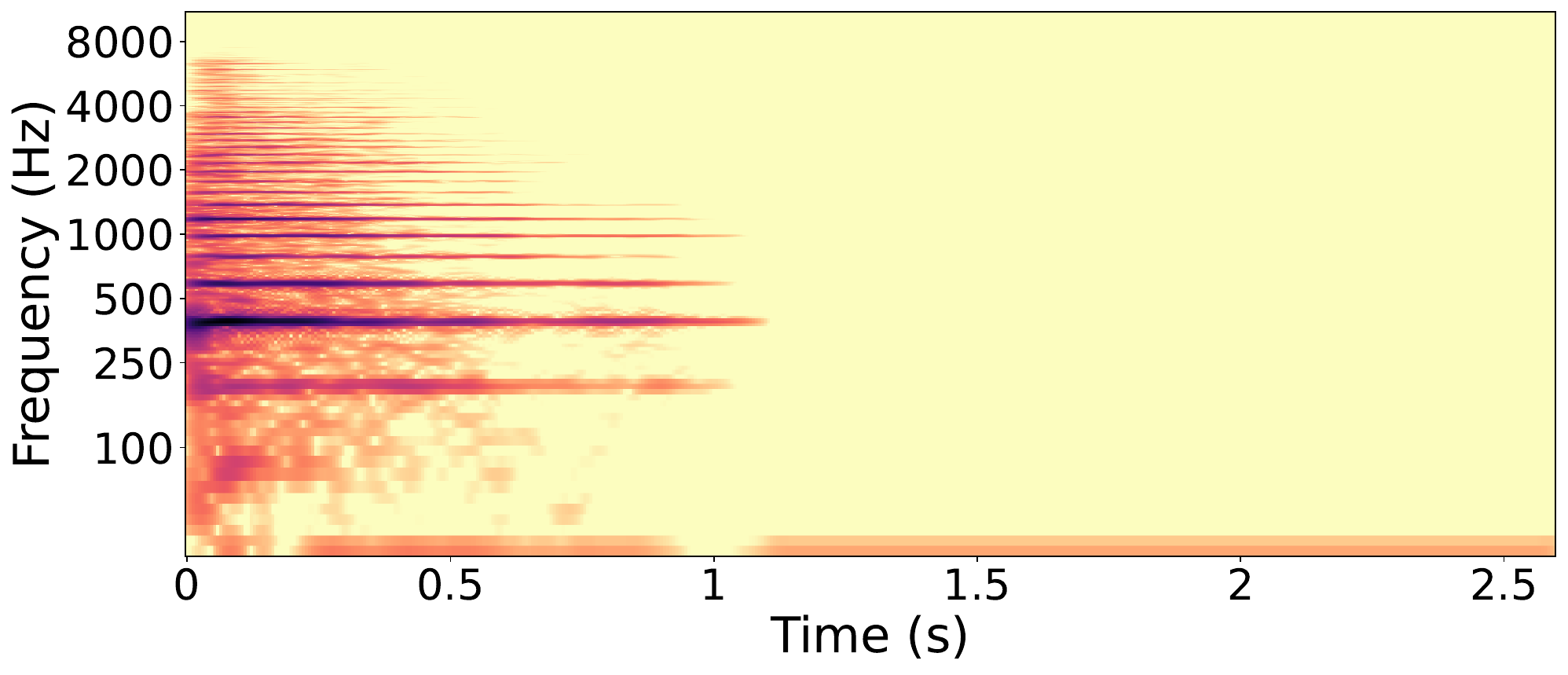}
        }
        \caption{Above: Spectrogram of the RWC cello sample with the lowest difference between predicted ``resonant-vibrant" value and ground truth. Below: Spectrogram of the synthesized cello sample with the highest difference between predicted ``resonant-vibrant" value and ground truth.}
        \label{fig:Synth Cello Spectrograms Resonant}
    \end{figure}
    
    We can also analyze more subtle defects by observing any timbre trait of any instrument. For instance, we can analyze the ``resonant/vibrant" trait of the cello. Its predicted TTP is presented in Figure \ref{fig:Synth Cello TTP}. The ``resonant/vibrant" ground truth value is 0.865, the highest value for this trait in the 31 instruments considered here, suggesting it is an essential timbral quality for the cello. Figure \ref{fig:Synth Cello Spectrograms Resonant} shows the spectrograms of a) the synthesized cello sample with the highest difference between predicted ``resonant-vibrant" value and ground truth, and b) the RWC cello sample with the lowest difference between predicted ``resonant/vibrant" value and ground truth. The prediction of this trait for the synthesized sample is 0.66. By observing the rapid amplitude decay of the synthesized sample's spectrogram and by listening to it, we notice that TokenSynth synthesized a plucked note sample whereas the RWC sample uses bowing technique. Difference in performance technique leads of course to significant timbre differences. The spectrogram of the RWC sample shows harmonics whose frequencies fluctuate slightly over time, plausibly creating a ``vibrant" perceptual quality related to the pronounced vibrato in the sample. In contrast, the synthesized sample’s spectrogram lacks these frequency oscillations in its harmonics.  This difference helps explain why the synthesized sample receives a lower prediction score for the ``resonant/vibrant" attribute.\footnote{The sounds discussed here as well as others can be listened to via the companion page.}

    Following this method, we can utilize the predicted TTPs to assess the synthesized instrument samples, identify shortcomings, and suggest new design strategies to potentially improve the performance of the synthesizer. 
    
	\section{Discussion}\label{sec:conclusion}

    In this paper, we propose a new machine listening task: Timbre Traits Profiles prediction. We 1) demonstrate the effectiveness of a simple learnable reweighting of deep neural audio embeddings and 2) exemplify the use of TTP prediction for assessing the performance of the musical sound synthesizer TokenSynth.
    
    The objective is to predict the timbres of instrument samples using TTP from Reymore \cite{Reymore2021characterizing}, which is constructed from human ratings, as ground truth. We introduce a deep neural timbre trait predictor called TTP-RANE composed of a pre-trained neural embedding and a shallow multi-layer perceptron, reweighting the neural embedding. To train our model, we compare four deep neural embeddings: CLAP (and its music-only version), MERT, and VGGish on samples from the RWC dataset. TTP-RANE performance is evaluated via Pearson Correlation during cross-evaluation, with the best result (0.66) achieved by a no-hidden-layer model trained on CLAP embeddings. We then train a global model called TTP-RANE-CLAP on all 31 instruments and apply it to assess the synthesis quality of TokenSynth, guided by CLAP embeddings, across three synthesis types: text-conditioned, audio-conditioned, and a hybrid conditioned by averaging text-audio embeddings. Audio-conditioned synthesis yields the most accurate TTPs predictions, and we observe that text conditioning added to audio conditioning does not improve the predictions. The ranking of synthesis quality based on the Mean Absolute Error (MAE) between predictions and ground truth aligns with the Fréchet Audio Distance (FAD) using the RWC dataset as a reference, indicating the TTP-RANE robustness in capturing distributional timbre information. Finally, we demonstrate how TTP prediction can be utilized to analyze and interpret synthesis defects, using the wood block and the cello as examples.
    
    From a technical point of view, future work will address 1) new deep neural architectures to improve our TTP prediction model and 2) new ways to condition or train musical audio synthesizers using respectively a TTP-RANE inspired conditioner or loss. Finally, while the TTP-RANE-CLAP model has been studied here for the sole evaluation of musical audio synthesizer, it may also be useful in music analysis. Reymore \cite{reymore2022modeling} suggests that TTPs may be used to identify when instruments are utilized in archetypical v. atypical or unexpected ways. TTP-RANE offers a way of accomplishing this computationally, by predicting timbre profiles for notes or excerpts from a recorded piece, and using MAE as a metric for typicality. This metric could be used in music analysis, both to support close readings and to test hypotheses through corpus studies. Additionally, a refined TTP prediction model could inform understanding of timbre semantics, potentially contributing to characterizing the nature of timbral interactions with other musical parameters, such as pitch, loudness, and articulation. Initial work has been done in this area \cite{reymore2023timbre, reymore2021variations}, but the enormous variation possible across musical parameters presents a significant challenge, as it cannot be practically modeled through human judgments.  TTP prediction could be thus used to map an instrument's timbral affordances across registers, dynamic ranges, and different playing techniques. Further testing could also determine whether TTP prediction is a promising approach to quantifying timbral qualities of instrumental blends. These timbral affordance maps would have practical applications in analysis, composition, and orchestration by offering access to average perceptual qualities of particular timbres without needing to acquire new sets of human judgments for each specific context. 
    
	\bibliography{bibliography}
	
\end{document}